\documentclass[
    aps,prl,twocolumn,
    superscriptaddress,
    longbibliography,
    floatfix,
    balancelastpage
]{revtex4-2}

\usepackage{amsmath}
\usepackage{amssymb}
\usepackage{braket}
\usepackage{graphicx}   
\usepackage{dcolumn}    
\usepackage{bm}         
\usepackage{siunitx}
\usepackage{calrsfs}
\DeclareMathAlphabet{\pazocal}{OMS}{zplm}{m}{n}
\usepackage{rotating}
\usepackage{color}
\usepackage{mathptmx}  
\usepackage{hyperref}
\hypersetup{hidelinks}
\newcommand{\figref}[1]{\hyperref[{#1}]{\textup{Fig.~\ref*{#1}}}}
\usepackage{ORCIDinREVTeX}
\usepackage{pdfpages}
\usepackage{pgffor}

\makeatletter
\AtBeginDocument{\let\LS@rot\@undefined}
\makeatother

\newcommand{\PT}{$\pazocal{PT}$}

\begin{document}
\title{Beyond the \PT-Broken Phase: Constant-Frequency Oscillation from Floquet Hybridization}

\author{Xiaoke Gao}
\orcid{0000-0002-3769-4451}
\affiliation{School of Electrical Engineering, Chongqing University, Chongqing 400044, China.}%

\author{Xiaoyin Sha}
\orcid{0009-0006-0180-2365}
\affiliation{School of Electrical Engineering, Xi'an Jiaotong University, Xi'an 710049, China.}%

\author{Xunyan Sun}
\orcid{0009-0001-2846-7070}
\affiliation{School of Electrical Engineering, China University of Mining and Technology, Xu Zhou 221116, China.}%

\author{Xianglin Hao}
\orcid{0000-0001-7149-0113}
\affiliation{Department of Electrical Engineering, City University of Hong Kong, Hong Kong, China}%

\author{Huaiqing Zhang}
\orcid{0000-0002-2631-4740}
\email[Email: ]{zhanghuaiqing@cqu.edu.cn}
\affiliation{School of Electrical Engineering, Chongqing University, Chongqing 400044, China.}%

\author{Tianyu Dong}
\orcid{0000-0003-4816-0073}
\email[Email: ]{tydong@mail.xjtu.edu.cn}
\affiliation{School of Electrical Engineering, Xi'an Jiaotong University, Xi'an 710049, China.}%

\date{September 30, 2026}

\begin{abstract}
    Constant-frequency oscillation is usually tied to the broken phase of a spatially coupled \PT-symmetric dimer, where the eigenfrequencies form a complex-conjugate pair with a fixed real part. This work shows that an analogous locking can instead come from Floquet hybridization in a synthetic frequency dimension. Periodic gain--loss modulation folds the spectrum into replicas spaced by the drive frequency $\Omega$. Coupling between neighboring replicas opens an additional real-part degeneracy at the edge of the Floquet Brillouin zone. The hybridized Floquet mode at $\omega_0\pm\Omega/2$ requires the minimum gain and is therefore selected for the self-oscillation. A Floquet \PT-symmetric circuit with a JFET time-varying negative resistance confirms the prediction. The observed constant-frequency oscillation contains two frequency components of comparable amplitude, separated by the modulation frequency, which identifies its origin from Floquet hybridization rather than ordinary \PT-symmetry breaking or dispersion flattening. More broadly, once a suitably symmetry-constrained spectral degeneracy is established, self-sustained frequency pinning can arise beyond the conventional spatial \PT-broken phase. 
\end{abstract}

\maketitle

\begin{figure}[!ht]
    \centering
    \includegraphics[width = 0.98\linewidth]{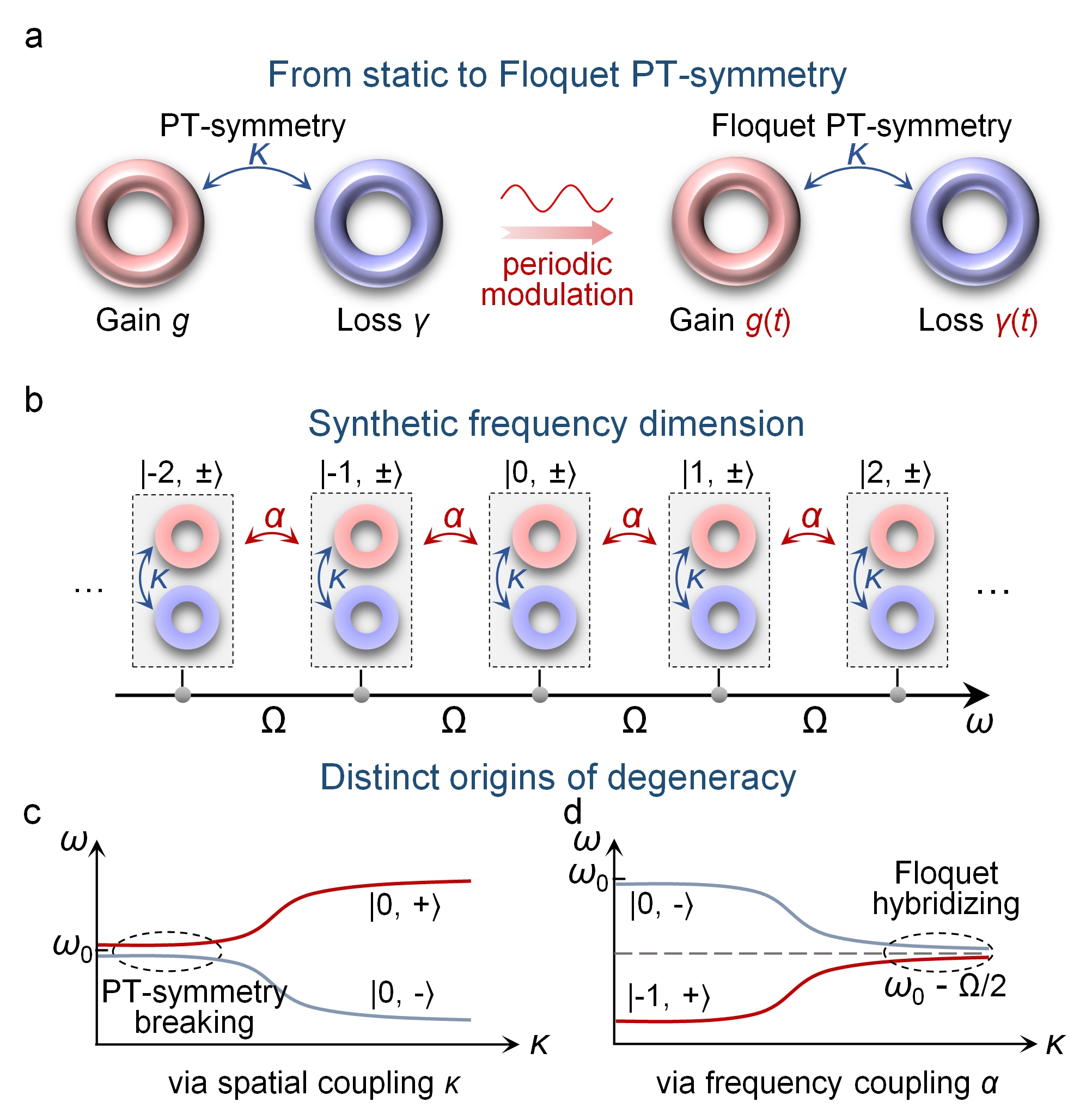}
    \caption{Schematic of static and Floquet \PT-symmetric dimers. (a) A static \PT-symmetric dimer becomes Floquet-driven under periodic gain--loss modulation. (b) Synthetic frequency lattice with spatial coupling $\kappa$ and frequency-dimension coupling $\alpha$. (c, d) Degeneracies from \PT-symmetry breaking by $\kappa$ and from Floquet hybridization by $\alpha$, respectively.}
    \label{fig:figure01}
\end{figure}

Symmetry fundamentally shapes the spectral properties of coupled systems, giving rise to characteristic phenomena such as eigenvalue coalescence, spectral bifurcation, and real-part degeneracy in the complex-frequency plane. A paradigmatic example is the \PT-symmetric dimer, consisting of two spatially coupled resonators with balanced gain and loss, as shown in \figref{fig:figure01}{(a)}. The underlying \PT-symmetry produces a distinctive spectral phase transition \cite{bender1998real,bai2026recent}: at a critical coupling strength, the two eigenfrequencies coalesce at an exceptional point (EP), whereas in the \PT-broken phase they form a complex-conjugate pair while retaining an invariant real part. This intimate correspondence between symmetry and spectral structure naturally raises a broader question: is such a spectral configuration unique to \PT-symmetry in spatially coupled systems, or can a different symmetry, acting through an alternative coupling dimension, produce analogous spectral bifurcations and degeneracies? If so, physical functionalities widely regarded as requiring EPs or the \PT-symmetry-broken phase --- such as EP-enhanced sensing \cite{chen2018generalized,chen2024inductor,yin2026exceptional,ma2026critical,chen2026nonlinear} and constant-frequency wireless power transfer \cite{assawaworrarit2017robust,assawaworrarit2020robust,kim2022wide,hao2023frequency,guo2024level,hao2025dispersive} --- could be extended beyond the conventional \PT-symmetric paradigm by exploiting new coupling dimensions and symmetries. Periodic temporal modulation offers a concrete setting in which to address this question.

In the Floquet \PT-symmetry dimer illustrated in \figref{fig:figure01}{(a)}, periodic gain--loss modulation introduces a distinct temporal symmetry of the modulation period $T = 2\pi/\Omega$. As in a photonic crystal, this residual symmetry folds the spectrum into Floquet replicas spaced by $\Omega$, which establish a synthetic frequency dimension alongside spatial coupling \cite{maamache2017pseudo,chitsazi2017experimental,li2020parity} in \figref{fig:figure01}{(b)}. Within the given Floquet sector the spatial coupling $\kappa$ hybridizes the branches $\ket{0,+}$ and $\ket{0,-}$; the Floquet-induced coupling $\alpha$ hybridizes $\ket{0,-}$ with $\ket{-1,+}$ on the neighboring replica. These two processes produce distinct degeneracies: a conventional \PT-symmetry-breaking degeneracy at $\omega_0$ [\figref{fig:figure01}{(c)}], and a Floquet-hybridizing degeneracy at $\omega_0 - \Omega/2$ [\figref{fig:figure01}{(d)}]. Such a coupling feature therefore generates new types of degeneracies and hybridized modes absent from static \PT-symmetric framework, and implies that constant-frequency oscillation ascribed to a \PT-broken phase can instead arise from inter-replica coupling and symmetry in the frequency dimension.

Electronic circuits realize these settings with independently tunable parameters, balanced gain and loss, and controlled temporal modulation \cite{sahin2025topolectrical,stegmaier2025topological,fernandez2026parity}. Because active circuits support self-oscillations, Floquet reconstruction of the quasi-frequency spectrum can be used directly for frequency generation and stabilization. Floquet \PT-symmetry systems have been demonstrated in acoustics \cite{tong2025observation} and photonics \cite{liu2024floquet}, but those experiments, and related theory \cite{blose2020floquet,park2022revealing,shi2024floquet} address spectra, eigenstates, and topology under external drive. In a \PT-symmetric circuit, the spectrum supplies only candidate modes. In applications such as wireless power transfer, a constant-frequency spectral branch is useful only if it is the mode that is ultimately sustained. The central question is therefore which reconstructed mode is selected in the self-oscillation. Mapping that correspondence is the prerequisite for circuit functionalities based on Floquet engineering.

\begin{figure}[!ht]
    \centering
    \includegraphics[width = 0.98\linewidth]{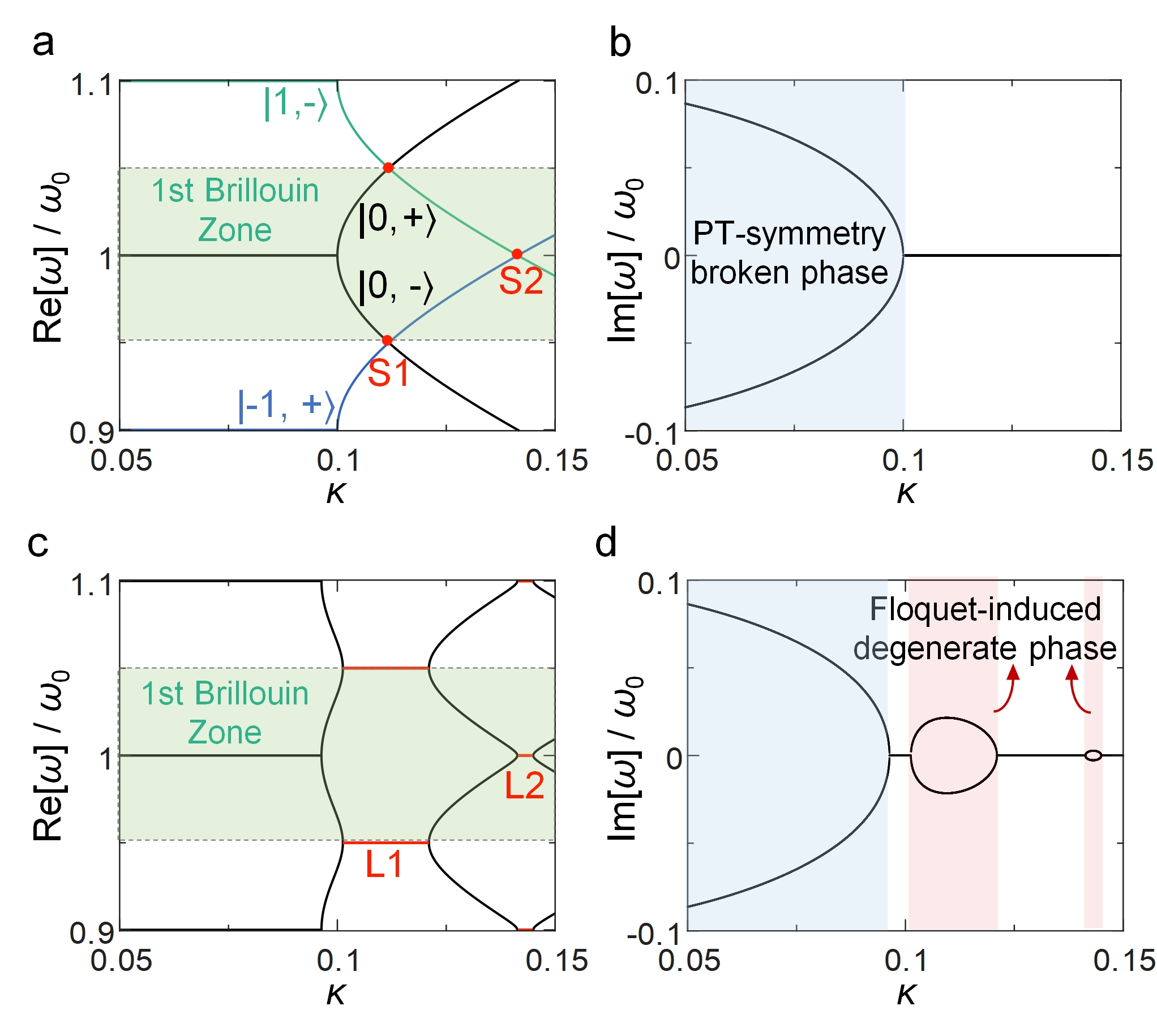}
    \caption{Floquet quasifrequency spectrum of the driven \PT-symmetric system for $g_0 = \gamma_0 = 0.1$ and $\Omega = 0.1\omega_0$. (a,b) Real and imaginary parts of the eigen-frequencies at vanishing modulation, $\alpha\simeq 0$. (c,d) Same for finite Floquet modulation, $\alpha = 0.2$. Green, blue, and red shadings mark the first Floquet Brillouin zone, the conventional \PT-broken phase, and the Floquet-induced degenerate phases, respectively.}
    \label{fig:figure02}
\end{figure}

We consider a non-Hermitian dimer with co-modulated gain $g(t) = g_0 \left(1 + \alpha\cos\Omega t\right)$ and loss $\gamma(t) = \gamma_0 \left(1 + \alpha\cos\Omega t\right)$, governed by the coupled-mode equations
\begin{equation}\label{eq:governingEquation}
    -\text{i}\frac{\text{d}}{\text{d}t}
    \begin{bmatrix}
        a_1 \\ a_2
    \end{bmatrix} =
    \begin{bmatrix}
        \omega_0 - \text{i}g(t) & -\kappa                      \\
        -\kappa                 & \omega_0 + \text{i}\gamma(t)
    \end{bmatrix}
    \begin{bmatrix}
        a_1 \\  a_2
    \end{bmatrix},
\end{equation}
where $\bm{a}(t) = [a_1, a_2]^\text{T}$, $\omega_0$ is the common bare resonance frequency, and $\kappa$ is the spatial coupling. Substituting the Floquet ansatz $\bm{a}(t) = \sum_{n\in\mathbb{Z}} \bm{u}_n \text{e}^{\text{i} (\omega - n\Omega) t}$, with $\omega$ the Floquet quasifrequency, we convert the time-periodic dynamics \eqref{eq:governingEquation} into the eigenvalue problem
\begin{equation}\label{eq:eigensystem}
    \bm{H}(\kappa)\bm{u} = \omega \bm{u}
\end{equation}
for the stacked amplitudes $\bm{u} = (\dots, \bm{u}_{-1}, \bm{u}_0, \bm{u}_1, \dots)^\text{T}$. The
Floquet matrix $\bm{H}(\kappa)$ is block tridiagonal, with blocks
\begin{equation}
    \bm{H}_{mn}(\kappa) = \left[H_0(\kappa)+n\Omega \mathbb{I}_2\right]\delta_{m,n} + V_{+}\delta_{m,n+1} + V_{-}\delta_{m,n-1},
\end{equation}
where $\delta_{m,n}$ is the Kronecker delta and $\mathbb{I}_2$ is the $2\times2$ identity matrix. In addition, we have
\begin{subequations}
    \begin{align}
        H_0(\kappa)   & =
        \begin{bmatrix}
            \omega_0-\mathrm{i}g_0 & -\kappa                     \\
            -\kappa                & \omega_0+\mathrm{i}\gamma_0
        \end{bmatrix}, \\
        V_{+} = V_{-} & =
        \begin{bmatrix}
            -\mathrm{i}\alpha g_0/2 & 0                           \\
            0                       & \mathrm{i}\alpha \gamma_0/2
        \end{bmatrix}.
    \end{align}
\end{subequations}
The quasifrequency spectrum $\omega = \omega(\kappa)$ follows from $\text{det}[\bm{H}(\kappa) - \omega \mathbb{I}] = 0$ and is shown in \figref{fig:figure02}. For vanishing modulation, $\alpha\simeq 0$ [\figref{fig:figure02}{(a,b)}], the spectrum comprises the \PT-symmetric branches $\ket{0, \pm}$ (black) together with the Floquet-replica branches $\ket{-1, +}$ (blue) and $\ket{1, -}$ (green). Because the inter-harmonic matrix element $\bra{0, -} V_{+} \ket{-1, +} = \pazocal{O}(\alpha)$ is negligible, neighboring replicas cross rather than hybridize, producing point degeneracies at the center and edge of the first Brillouin zone $[\omega_0 - \Omega/2, \omega_0 + \Omega/2]$ (green shading). Points S1 and S2 mark the crossings $\left\{\ket{0, -}, \ket{-1, +}\right\}$ and $\left\{\ket{1, -}, \ket{-1, +}\right\}$, respectively. The imaginary spectrum in \figref{fig:figure02}{(b)} shows the conventional \PT-broken phase at weak coupling (blue shading). At $\alpha = 0.2$, inter-replica coupling between neighboring Floquet harmonics hybridizes the crossing branches. The isolated degeneracies S1 and S2 therefore open into finite intervals of real-part degeneracy L1 and L2 [\figref{fig:figure02}{(c)}], while the imaginary parts split over the same range of $\kappa$ [red shading in \figref{fig:figure02}{(d)}]. From a spectral perspective, the conventional \PT-broken phase and Floquet-induced degenerate phase are quite similar, despite arising from spatial coupling $\kappa$ and frequency-domain coupling $\alpha$, respectively.

\begin{figure*}[!hbt]
    \centering
    \includegraphics[width = 0.98\linewidth]{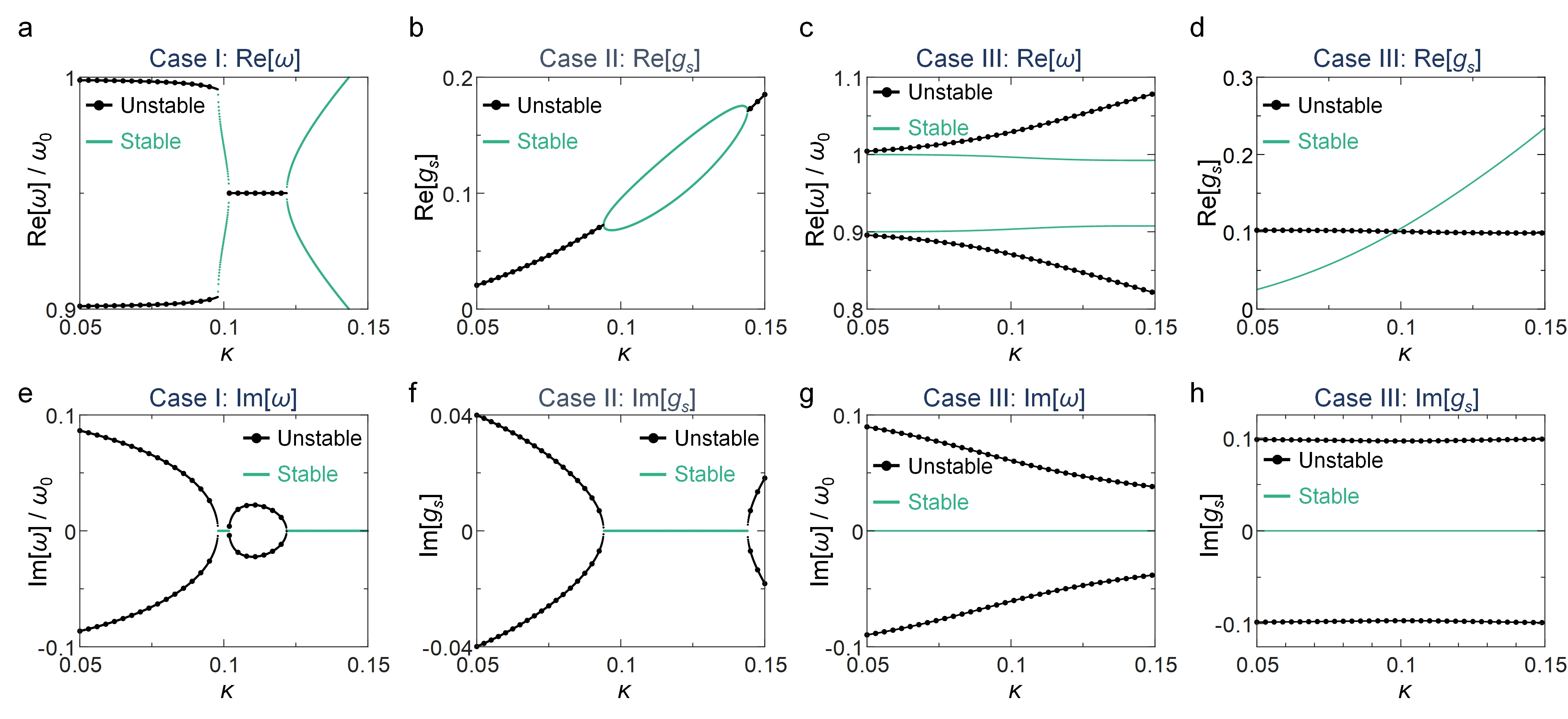}
    \caption{Solution classes of the reduced Floquet problem $\bm{H}_{\text{sub}}(\kappa)$. (a, e) Case I: Floquet quasifrequency $\omega$ at $g_s = \gamma_0$. (b, f) Case II: static gain $g_s$ at $\omega = \omega_0 - \Omega/2$. (c,g) Quasifrequency $\omega$ and (d, h) static gain $g_s$ for case III. Top (bottom) row: real (imaginary) parts. Green (black): stable (unstable) branches. Here, parameters are: $\gamma_0 = 0.1$, $\alpha = 0.2$, and $\Omega = 0.1\omega_0$.}
    \label{fig:figure03}
\end{figure*}
Since frequency degeneracy enables constant-frequency modes in the conventional \PT-broken phase \cite{assawaworrarit2017robust}, we ask whether the Floquet-induced degenerate phase does the same. Near the degenerated band L1 we truncate the full Floquet matrix to the four-dimensional subspace spanned by $\{\ket{0,\pm}, \ket{-1,\pm}\}$,
\begin{equation} \label{eq:ResHamiltonian}
    \bm{H}_\text{sub}(\kappa) = \begin{bmatrix}
        H_0(\kappa)-\Omega \mathbb{I}_2 & V_{-} \\ V_{+} & H_0(\kappa)
    \end{bmatrix}.
\end{equation}
Treating the static gain as an unknown $g_s$ rather than the fixed parameter $g_0$, the Floquet quasifrequencies of $\bm{H}_\text{sub}(\kappa)$ satisfy
\begin{equation}\label{eq:Eigen}
    \text{det}[\bm{H}_\text{sub}(\kappa,g_s) - \omega\mathbb{I}_4] = 0,
\end{equation}
whose imaginary part factors as
\begin{equation} \label{eq:ImEigen}
    (g_s - \gamma_0) \cdot \Delta\omega \cdot (4\Delta\omega^2 + 4 g_s\gamma_0 - g_s \gamma_0 \alpha^2 - 4\kappa^2 - \Omega^2) = 0,
\end{equation}
with $\omega = \omega_0 - \Omega/2 + \Delta\omega$. The solution of \eqref{eq:ImEigen} define the branches in \figref{fig:figure03}. In case I, $g_s = \gamma_0$, the quasifrequencies are the real solutions of \eqref{eq:Eigen} [\figref{fig:figure03}{(a, e)}]. Balanceed gain and loss lock $\text{Re}[\omega]$ in both the conventional \PT-broken phase and the Floquet-induced degenerate phase, but a nonzero $\text{Im}[\omega]$ precludes steady oscillations. Stable modes therefore exist only on the nondegenerate branches, as in the exact \PT-symmetric phase. In case II, $\Delta\omega = 0$, the real part of \eqref{eq:Eigen} yields a real $g_s(\kappa)$ over a finite interval of coupling [\figref{fig:figure03}{(b, f)}]. The quasifrequency is pinned at $\omega_0 - \Omega/2$, a Floquet-induced constant-frequency regime analogous to conventional \PT-broken phases. In case III, $4\Delta\omega^2 + 4 g_s\gamma_0 - g_s \gamma_0 \alpha^2 - 4\kappa^2 - \Omega^2 = 0$ together with the real part of $\det[\bm{H}_\text{sub} - \omega\mathbb{I}_4] = 0$ determines the $\omega(\kappa)$ and $g_s(\kappa)$ [\figref{fig:figure03}{(c, g)} and \figref{fig:figure03}{(d, h)}]. Here, only the two green branches are physical: two distinct real quasifrequencies supported by the same real static gain. At weak coupling, both quasifrequencies are nearly $\kappa$-independent as in the conventional \PT-broken phase. Cases II and III reveal two distinct routes to constant-frequency oscillation, demonstrating that this behavior extends beyond the \PT-broken phase associated with spatial coupling.

\begin{figure}[!ht]
    \centering
    \includegraphics[width=0.98\linewidth]{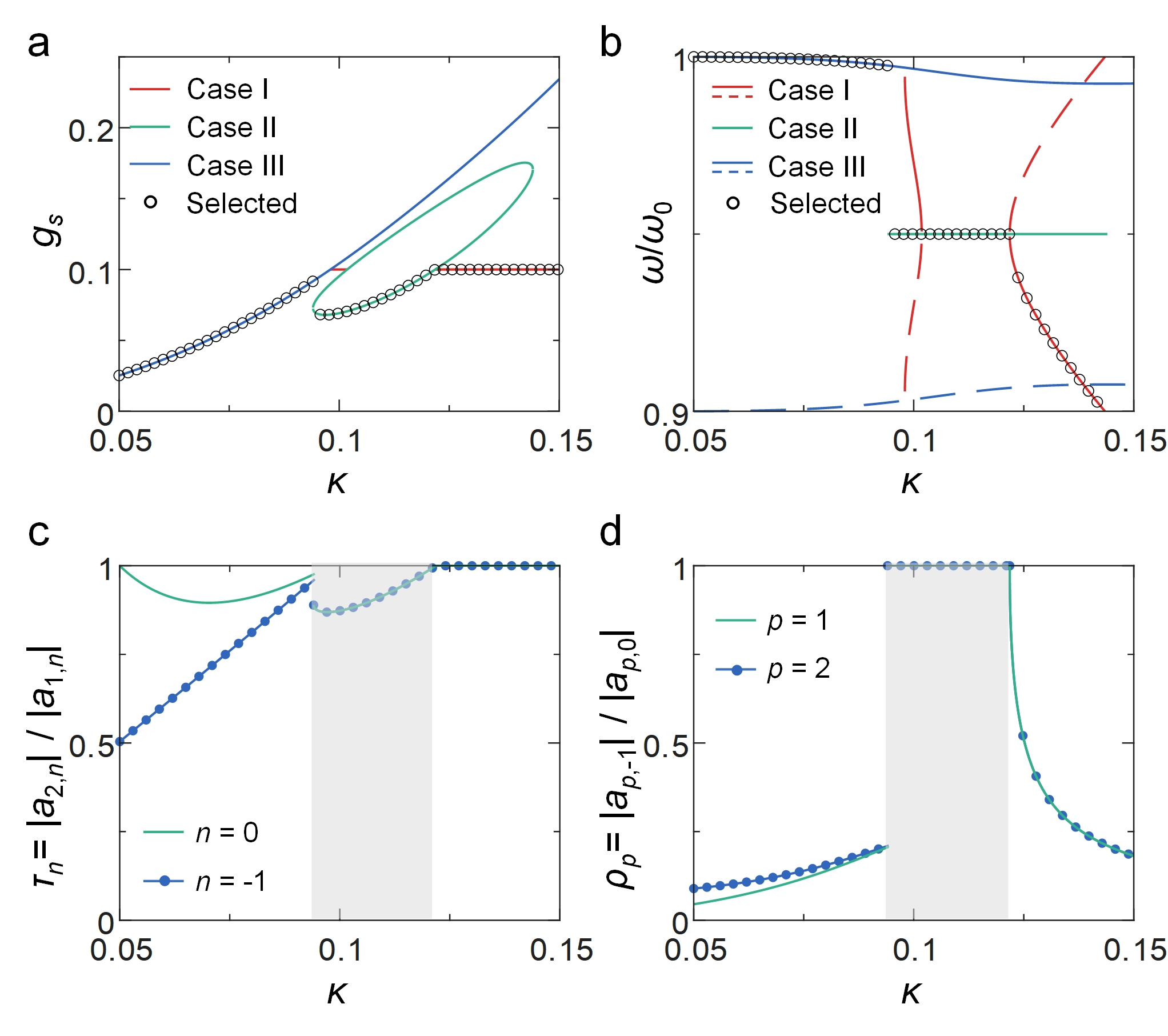}
    \caption{Minimum-gain selection and modal composition of stable Floquet modes. (a) Required static gain $g_s$ and (b) Floquet quasifrequency $\omega$ of the candidate branches versus $\kappa$. Red, green, and blue curves denote cases I -- III, respectively; circles mark the selected minimum-gain solutions for excitation. (c) Inter-resonator modal transmission $\tau_n = |a_{2,n}|/|a_{1,n}|$ for the harmonics $n = 0, -1$. (d) Inter-harmonic ratio $\rho_p = |a_{p,-1}|/|a_{p,0}|$ in the gain ($p = 1$) and loss ($p = 2$) resonators. Gray shading marks the coupling range in which the Floquet-induced case II branch is selected.}
    \label{fig:figure04}
\end{figure}

The analysis above identifies candidate constant-frequency modes but does not determine which of them is sustained in a self-oscillating circuit. Following the minimum-gain principle \cite{assawaworrarit2017robust}, we rank the branches in \figref{fig:figure03} by their required static gain $g_s$ [\figref{fig:figure04}{(a)}]. At each $\kappa$, the lowest-gain solution (circles) is selected, and the winning cases switches among the three cases as $\kappa$ varies. In particular, the Floquet-induced constant-frequency branch (case II) is the minimum-gain solution over a finite interval inside the Floquet-induced degenerate phase, and is therefore directly excitable and observable. The corresponding quasifrequencies are shown in \figref{fig:figure04}{(b)}. Branches that differ by integer multiples of $\Omega$ are equivalent harmonic Floquet replicas, so only the $n=0$ representatives enter the minimum-gain comparison; dashed replicas of cases I and III are omitted. \figref{fig:figure04}{(c)} and \figref{fig:figure04}{(d)} exhibit the modal compositions of the selected branches, where $|a_{p,n}|$ denotes the amplitude of the $n$th Floquet harmonic component in the gain ($p = 1$) or loss ($p = 2$) resonator. \figref{fig:figure04}{(c)} shows the inter-resonator transmission $\tau_n = |a_{2,n}/a_{1,n}|$ between the gain and loss resonators for each Floquet harmonic. The Floquet-induced constant-frequency oscillation (shaded region) occurs at larger $\kappa$ than the conventional \PT-broken phase, and the stronger spatial inter-resonator coupling produces only a modest rise in modal transmission $\tau_n$. Throughout most of this interval $\tau_n < 1$, so near-unity modal transmission is not obtained. \figref{fig:figure04}{(d)} shows the inter-harmonic amplitude ratio $\rho_p = |a_{p,-1}/a_{p,0}|$. On the selected case-II interval, $\rho_1 = \rho_2 = 1$: the $n = 0$ and $n = -1$ harmonics have equal weight in both resonators. These components sit at the physical frequencies $0.95\omega_0$ and $1.05\omega_0$, separated by $\Omega = 0.1\omega_0$, yet belong to a single Floquet mode of quasifrequency $0.95\omega_0$. The coexistence of two frequency components with comparable amplitudes contrasts with ordinary gain-saturated self-oscillation, which typically selects one dominant frequency and suppresses the rest. This spectral feature provides direct evidence that the case-II oscillation arises from hybridization of the resonant channels $\ket{0,-}$ and $\ket{-1,+}$ at the edge of Floquet Brillouin zone, rather than from conventional \PT-symmetry breaking or from dispersion flattening.

\begin{figure*}[!ht]
    \centering
    \includegraphics[width = 0.96\linewidth]{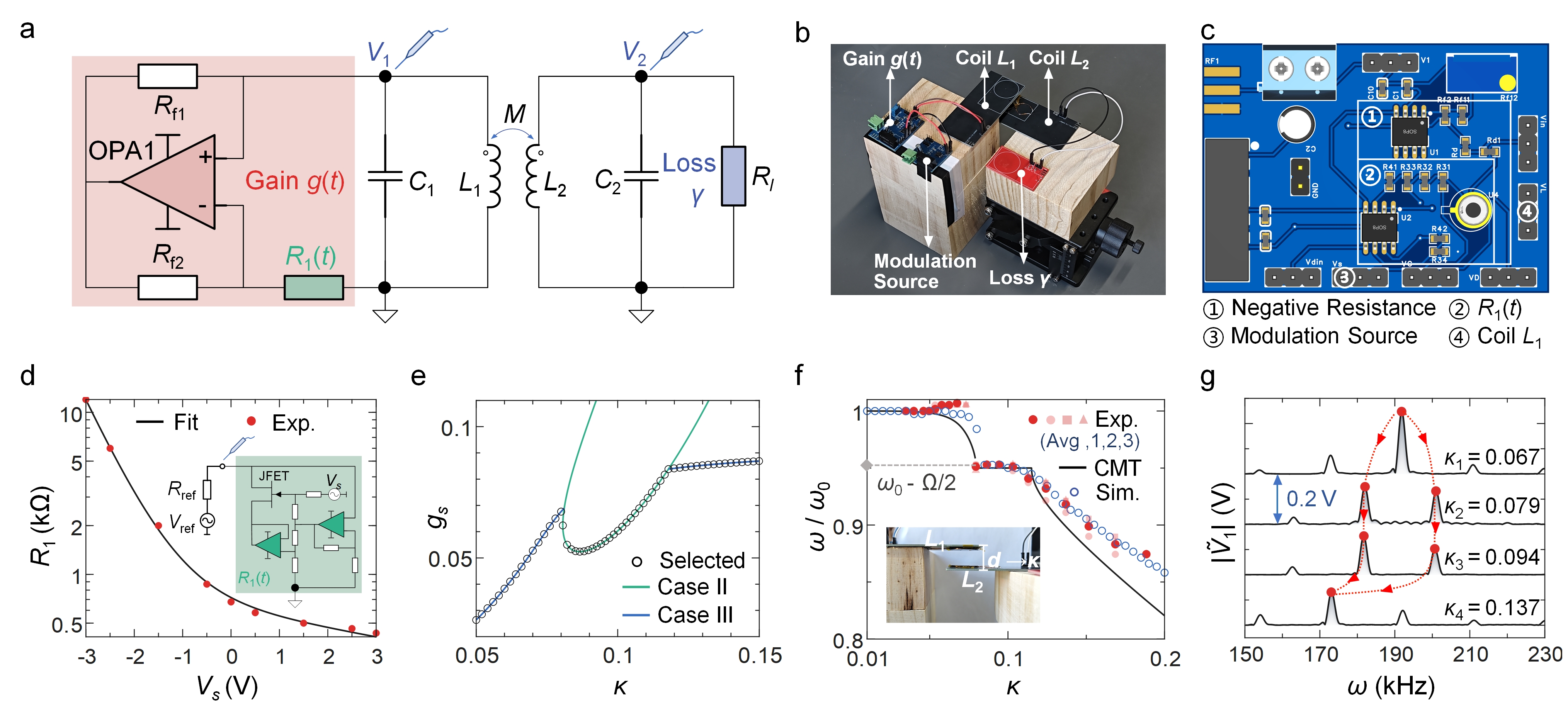}
    \caption{Experimental realization of Floquet-induced constant-frequency oscillation. (a) Equivalent circuit: two inductively coupled LC resonators with modulated gain. (b) Photograph of the setup. (c) Printed circuit board of the gain resonator, with an operational-amplifier negative resistor and a JFET implementing $R_1(t)$. (d) Measured $R_1$ versus modulated voltage $V_\text{s}$. The inset shows the divider circuit used to extract $R_1$ and the theoretical JFET model. (e) Required static gain $g_s$ of the candidate branches, with circles marking the selected minimum-gain solutions. (f) Oscillation frequency versus $\kappa$ from coupled-mode theory (CMT), circuit simulation, and experiment. Pale pink symbols show three independent runs; red circles indicate their mean. The gray line marks the locked frequency $0.95\omega_0$. The inset shows coupling tuned by coil separation. (g) Measured voltage spectra at $\kappa = 0.067, 0.079, 0.094$ and $0.137$. Red markers identify the dominant frequency components, and neighboring traces are offset by $0.2~\si{\volt}$. The circuit parameters and methods are given in the Supplemental Material.}
    \label{fig:figure05}
\end{figure*}

We test the theory on a Floquet-driven \PT-symmetric circuit of two inductively coupled LC resonators [\figref{fig:figure05}{(a)}]. Gain is supplied by a negative resistance and loss $\gamma$ by a passive resistor. Periodic gain modulation is realized with a time-varying feedback resistor $R_1(t)$, so that $g(t) \propto R_{f2}/[R_{f1}R_1(t)]$. The experimental platform shown in \figref{fig:figure05}{(b)} comprises five functional modules: a modulation source, a time-varying negative resistor, and coil $L_1$ on the gain side; coil $L_2$ and a passive resistor on the loss side. \figref{fig:figure05}{(c)} shows the gain-side board: an operational-amplifier (OPA) negative resistor in series with $R_1(t)$, implemented by a junction field-effect transistor (JFET) biased in its ohmic region. The modulation voltage $V_\text{s}$ sets the depletion width and hence the drain--source resistance. Measured $R_1(V_s)$ for the assembled module is plotted in \figref{fig:figure05}{(d)}. Over $V_\text{s} \in [-2, 2]~\si{\volt}$, $R_1$ varies by about $1.5~\si{\kilo\ohm}$, enough for deep gain modulation. The inset gives the measurement circuit: the module is placed in series with a known resistor $R_\text{ref}$ and $R_1$ is obtained from the voltage divider.

The circuit adopts $L_1 = L_2 = 46~\si{\micro\henry}$ and $C_1 = C_2 = 15~\si{\nano\farad}$, so $f_0 = 191.6~\si{\kilo\hertz}$. Loss is set to $\gamma_0 = 0.095$ by $R_l = 300~\si{\ohm}$, and the modulation frequency is $\Omega = 0.1\omega_0$. For $V_\text{s}(t) = -0.1 + 2\cos(\Omega t)~\si{\volt}$, the calibrated $R_1(V_\text{s})$ relation in \figref{fig:figure05}{(d)} gives the time-dependent gain $g(t)$. Fourier decomposition of $g(t)$ yields an effective modulation depth of $\alpha = 2|g_{\pm 1}|/g_0 = 0.763$. For practical implementation, only the gain resonator is modulated. This changes $V_{\pm}$ but leaves the discrete time-translate symmetry intact (Supplemental Material). The resulting required-gain branches are shown in \figref{fig:figure05}{(e)}: Case I collapses onto a subbranch of case III, while case II remains the minimum-gain branch over a finite interval of $\kappa$. \figref{fig:figure05}{(f)} shows the corresponding quasifrequency, with $\kappa$ set by the coil spacing $d$ (inset). At weak coupling, coupled-mode theory, circuit simulation, and experiment all recover the conventional constant-frequency oscillation at $\omega = \omega_0$ in the \PT-broken phase. At larger $\kappa$, all three reveal an additional constant-frequency interval at $\omega = \omega_0 - \Omega/2 = 0.95\omega_0$. Still stronger coupling keeps experiment and circuit simulation in agreement, while the prediction from coupled-mode theory drifts as its weak-coupling assumption fails. The spectrum of $V_1$ in \figref{fig:figure05}{(g)} follows the same three stages as \figref{fig:figure05}{(f)}. Before entering the constant-frequency region ($\kappa_1$), a single peak near $\omega_0$ marks conventional oscillation. Inside the frequency pinning region ($\kappa_2$ and $\kappa_3$), two comparable peaks appear at $\omega_0\pm\Omega/2$. They are separated by $\Omega$ in physical frequency but share the same quasifrequency, as predicted for Floquet hybridization. Beyond this interval ($\kappa_4$), the spectrum is again dominated by a single peak.

In conclusion, constant-frequency self-oscillation is not unique to the static \PT-broken phase. Periodic modulation hybridizes neighboring Floquet replicas, turns their isolated crossings into finite degeneracies, and thereby produces additional locked constant-frequency modes. A JFET-modulated circuit realizes these modes over the predicted range of $\kappa$. The oscillation carries two comparable tones separated by $\Omega$, identifying its origin from Floquet hybridization rather than conventional \PT-broken phase. This work reveals that coupling in different dimensions can produce analogous spectral degeneracies under distinct symmetry constraints, thereby giving rise to the similar frequency-pinning behavior.

\textit{Acknowledgments} --- T.~D. and H.~Z. acknowledge the support of National Natural Science Foundation of China (NSFC) under grant nos. 52577015 and U22B2095, respectively.

\textit{Data availability} --- There are no publicly available research data or software supporting this manuscript. Requests for further information or data should be sent to the authors.

\bibliography{main}

\clearpage
\onecolumngrid


\section*{Supplementary material}

# Supplemental Materials:
# Beyond the $\mathcal{PT}$-Broken Phase: Constant-Frequency Oscillation from Floquet Hybridization


Xiaoke Gao [1] Xiaoyin Sha [2] Xunyan Sun [3] Xianglin Hao [4] Huaiqing Zhang [1,*] and Tianyu Dong [2,†]

[1] *School of Electrical Engineering, Chongqing University, Chongqing 400044, China.*
[2] *School of Electrical Engineering, Xi'an Jiaotong University, Xi'an 710049, China.*
[3] *School of Electrical Engineering, China University of Mining and Technology, Xu Zhou 221116, China.*
[4] *Department of Electrical Engineering, City University of Hong Kong, Hong Kong, China*




## I. THEORY OF THE FLOQUET $\mathcal{PT}$-SYMMETRY SYSTEM

The dynamics of a non-Hermitian dimer with periodic gain $g(t)$ and loss $\gamma(t)$ can be described by

$$-\mathrm{i}\frac{\mathrm{d}}{\mathrm{d}t}\begin{pmatrix} a_1 \\ a_2 \end{pmatrix} = \begin{pmatrix} \omega_0 - \mathrm{i}g(t) & -\kappa \\ -\kappa & \omega_0 + \mathrm{i}\gamma(t) \end{pmatrix}\begin{pmatrix} a_1 \\ a_2 \end{pmatrix}, \tag{1}$$

where $a_{1,2}$ denote the modal amplitudes, $\omega_0$ is the common resonance frequency of the two resonators, and $\kappa$ is the coupling coefficient. In the main text, we take sinusoidal modulation $g(t) = g_0(1+\alpha\cos\Omega t)$ and $\gamma(t) = \gamma_0(1+\alpha\cos\Omega t)$ as a minimal and experimentally accessible form of periodic driving. Because a general periodic waveform can be decomposed into harmonic components, this single-harmonic case isolates the simplest frequency-domain coupling mechanism and provides a basis for treating more general modulation waveforms. Writing $\boldsymbol{a}(t) = [a_1, a_2]^{\mathrm{T}}$, (1) can be recast as $-\mathrm{i}\dot{\boldsymbol{a}}(t) = H(t)\boldsymbol{a}(t)$. By Floquet's theorem, solutions take the form $\boldsymbol{a}(t) = \boldsymbol{u}(t)\mathrm{e}^{\mathrm{i}\omega t}$, where $\omega$ denotes the Floquet quasifrequency and the periodic part $\boldsymbol{u}(t)$ has a period $T = 2\pi/\Omega$. Consequently, we have

$$-\mathrm{i}\dot{\boldsymbol{u}}(t) = [H(t) - \omega\mathbb{I}_2]\,\boldsymbol{u}(t), \tag{2}$$

where $\mathbb{I}_2$ denotes the $2\times 2$ identity matrix. With $\boldsymbol{u}(t) = \sum_n \boldsymbol{u}_n \mathrm{e}^{-\mathrm{i}n\Omega t}$ and $H(t) = H_0 + V_+\mathrm{e}^{-\mathrm{i}\Omega t} + V_-\mathrm{e}^{\mathrm{i}\Omega t}$, (2) is recast as

$$\omega\begin{pmatrix} \vdots \\ \boldsymbol{u}_{-1} \\ \boldsymbol{u}_0 \\ \boldsymbol{u}_1 \\ \vdots \end{pmatrix} = \boldsymbol{H}(\kappa)\begin{pmatrix} \vdots \\ \boldsymbol{u}_{-1} \\ \boldsymbol{u}_0 \\ \boldsymbol{u}_1 \\ \vdots \end{pmatrix} = \begin{pmatrix} \ddots & & & & \iddots \\ & H_0 - \Omega\mathbb{I}_2 & V_- & \mathbf{0} & \\ & V_+ & H_0 & V_- & \\ & \mathbf{0} & V_+ & H_0 + \Omega\mathbb{I}_2 & \\ \iddots & & & & \ddots \end{pmatrix}\begin{pmatrix} \vdots \\ \boldsymbol{u}_{-1} \\ \boldsymbol{u}_0 \\ \boldsymbol{u}_1 \\ \vdots \end{pmatrix}. \tag{3a}$$

where

$$H_0 = \begin{pmatrix} \omega_0 - \mathrm{i}g_0 & -\kappa \\ -\kappa & \omega_0 + \mathrm{i}\gamma_0 \end{pmatrix}, \qquad V_+ = V_- = \begin{pmatrix} -\mathrm{i}\alpha g_0/2 & 0 \\ 0 & \mathrm{i}\alpha\gamma_0/2 \end{pmatrix}. \tag{3b}$$

The relationship between the Floquet quasifrequency $\omega$ and the coupling coefficient $\kappa$ follows the characteristic equation

$$\det[\boldsymbol{H}(\kappa) - \omega\mathbb{I}_{2N}] = 0, \tag{4}$$

where $N$ is the truncating order of the infinite Floquet matrix $\boldsymbol{H}$.

## II. ANALYSIS OF FLOQUET DRIVING APPLIED ONLY TO THE GAIN RESONATOR

When Floquet driving is applied to the gain alone, $V_\pm$ reduces to

$$V_+ = V_- = \begin{pmatrix} -\mathrm{i}\alpha g_0/2 & 0 \\ 0 & 0 \end{pmatrix}. \tag{5}$$


* Email: zhanghuaiqing@cqu.edu.cn
† Email: tydong@mail.xjtu.edu.cn

The block-diagonal structure of the corresponding Floquet matrix $\boldsymbol{H}(\kappa)$ is nevertheless unchanged. Solving the Floquet eigenvalue problem as a function of $\kappa$ yields the complex-frequency spectra in Fig. S1. Panels (a,b) correspond to gain-only modulation and panels (c,d) to simultaneous gain and loss modulation; within each pair, the first panel shows Re[$\omega$, and the second shows Im[$\omega$]. Although gain-only modulation breaks the instantaneous balance between the gain and loss, the equality of the static components, $g_0 = \gamma_0$, preserves the conventional $\mathcal{PT}$-broken phase. Both configurations also exhibit Floquet-induced real-frequency degeneracies near the Brillouin-zone edges, $\omega = \omega_0 \pm \Omega/2$. Relative to simultaneous gain–loss modulation, however, gain-only driving produces a narrower flattened branch in Re[$\omega$] and a shallower splitting depth in Im[$\omega$], indicating a weaker effective Floquet coupling.

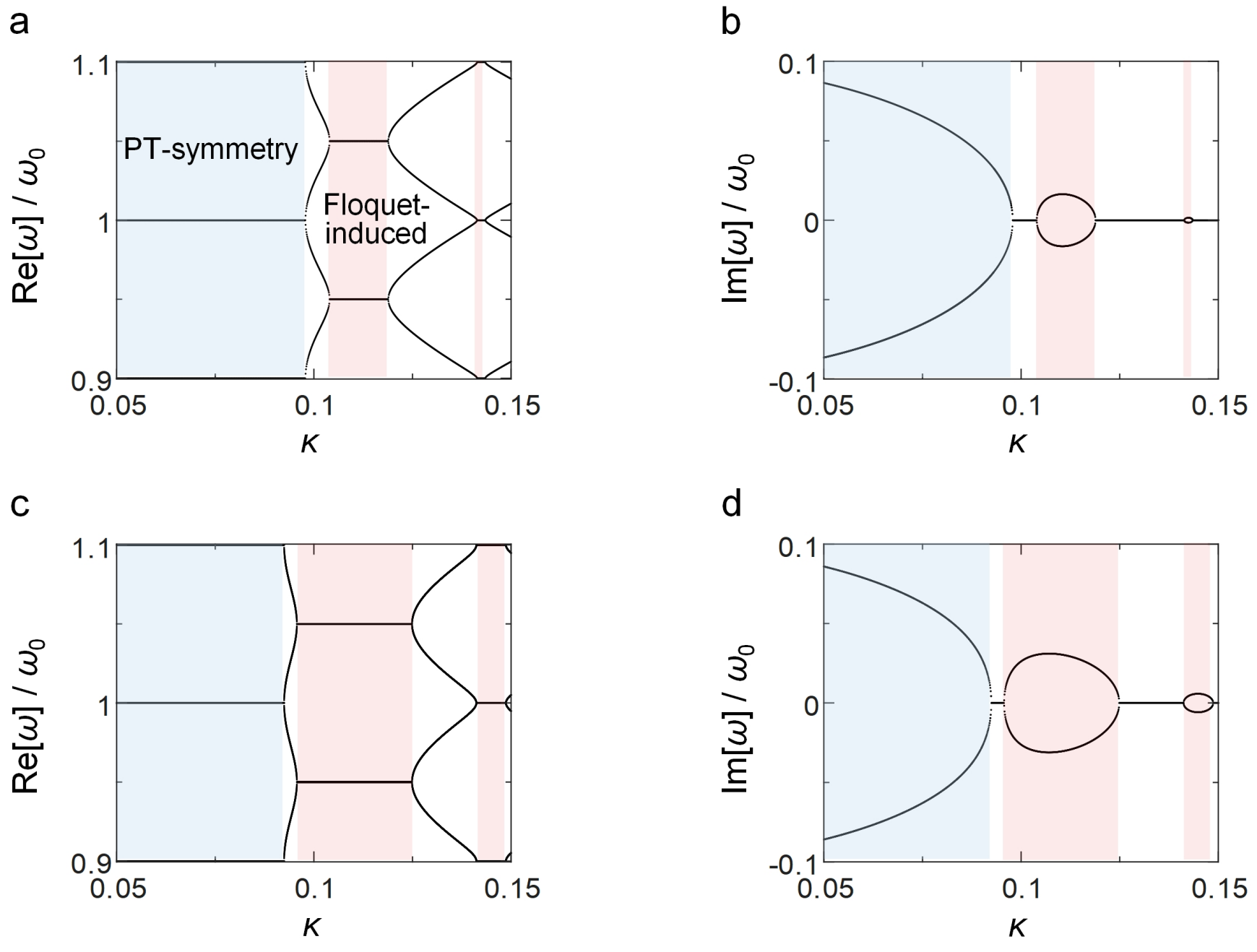


FIG. S1. Complex-frequency spectra for different Floquet-modulation configurations. (a,b) Gain-only modulation and (c,d) simultaneous gain and loss modulation. Panels (a,c) and (b,d) show the real and imaginary parts of the spectrum, respectively. Blue and red shaded regions mark the conventional $\mathcal{PT}$-symmetry-broken and Floquet-induced degenerate phases, respectively. The parameters are $g_0 = \gamma_0 = 0.1$, $\alpha = 0.3$, and $\Omega = 0.1\omega_0$.

To identify the modes that can be excited under gain-only modulation, we follow the procedure of the main text and truncate the Floquet matrix to the four-dimensional subspace spanned by $\{|0,\pm\rangle, |-1,\pm\rangle\}$,

$$\boldsymbol{H}_{\text{res}}(g_s,\kappa) = \begin{pmatrix} \omega_0 - \mathrm{i}g_s - \Omega & -\kappa & -\mathrm{i}\alpha g_s/2 & 0 \\ -\kappa & \omega_0 + \mathrm{i}\gamma_0 - \Omega & 0 & 0 \\ -\mathrm{i}\alpha g_s/2 & 0 & \omega_0 - \mathrm{i}g_s & -\kappa \\ 0 & 0 & -\kappa & \omega_0 + \mathrm{i}\gamma_0 \end{pmatrix}. \tag{6}$$

Here, $g_s$ is the static gain required to sustain the steady state at a given coupling strength $\kappa$. For prescribed $g_s$ and $\kappa$, the eigenfrequencies satisfy

$$\det[\boldsymbol{H}_{\text{res}}(g_s,\kappa) - \omega\mathbb{I}_4] = 0. \tag{7}$$

The imaginary part of (7) reduces to

$$(\omega - \omega_0 + \Omega/2)\cdot\left\{(g_s - \gamma_0)\cdot[4g_s\gamma_0 + 4(\omega - \omega_0 + \Omega/2)^2 - \Omega^2 - 4\kappa^2] - \alpha^2 g_s^2\gamma_0\right\} = 0. \tag{8}$$

Because the loss $\gamma_0$ is not modulated in the gain-only configuration, case I of the main text, $g_s = \gamma_0$, is no longer satisfied. Case II, $\omega = \omega_0 - \Omega/2$, remains valid, so a constant-frequency branch can still appear. For case III, $\omega$ is first obtained as a function of $g_s$ by solving the imaginary part of (7) subject to $\omega \neq \omega_0 - \Omega/2$. Substituting $\omega(g_s)$ into the real part of (7) then determines $g_s$ and the corresponding eigenfrequency $\omega$. Although the condition of case I is no longer satisfied independently, the associated branches are not eliminated: they are absorbed into the solution branch of case III, as shown in Fig. S2.

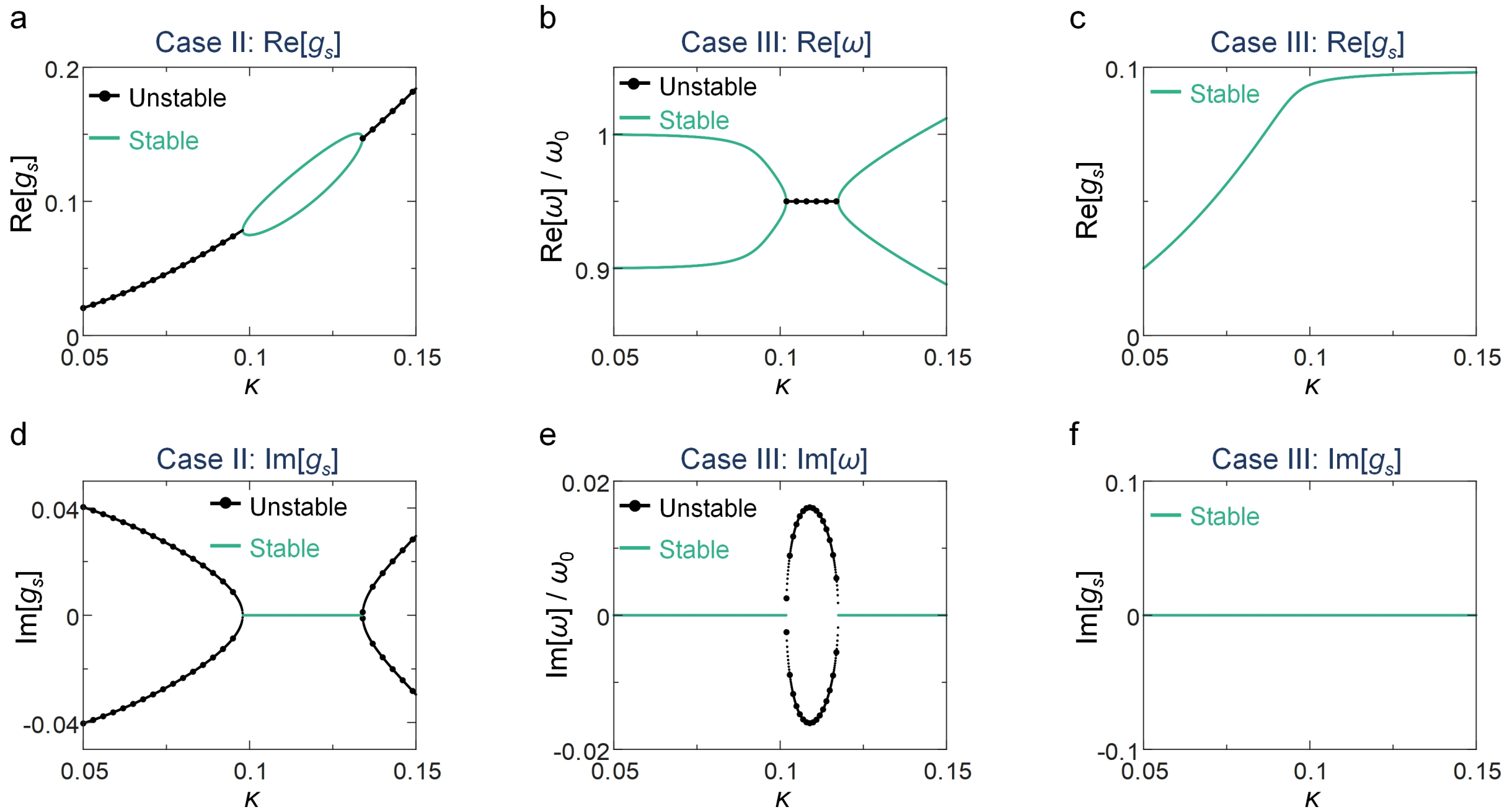


FIG. S2. Minimum-gain analysis for gain-only Floquet modulation. (a, d) Case II: static gain $g_s$ at $\omega = \omega_0 - \Omega/2$. (b, e) Quasifrequency and (c, f) required static gain for the case III branch selected by the stability and minimum-gain criteria. Upper and lower panels show the real and imaginary parts, respectively. Green curves denote stable modes and black curves unstable solutions. Parameters are the same as in Fig. S1.

## III. LTSPICE MODELING OF THE FLOQUET $\mathcal{PT}$-SYMMETRIC CIRCUIT

### A. Circuit realization

Applying Kirchhoff's current law to the circuit in Fig. S3(a), which is the same as Fig. 5(a) of the main text, we can obtain

$$I_{L_1} + \frac{V_1}{R_g(t)} + C_1\frac{\mathrm{d}V_1}{\mathrm{d}t} = 0, \tag{9a}$$

$$I_{L_2} + \frac{V_2}{R_l} + C_2\frac{\mathrm{d}V_2}{\mathrm{d}t} = 0. \tag{9b}$$

The voltages and currents of the mutually coupled inductors are related by

$$\begin{pmatrix} I_{L_1} \\ I_{L_2} \end{pmatrix} = \begin{pmatrix} L_1 & M \\ M & L_2 \end{pmatrix}^{-1} \cdot \begin{pmatrix} \int_0^t V_1(\tau)\,\mathrm{d}\tau \\ \int_0^t V_2(\tau)\,\mathrm{d}\tau \end{pmatrix}. \tag{10}$$

Eliminating $I_{L_1}$ and $I_{L_2}$ between (9) and (10) yields

$$C_1\frac{\mathrm{d}V_1}{\mathrm{d}t} + \frac{V_1}{R_g(t)} - \frac{L_2}{M^2 - L_1L_2}\cdot\int_0^t V_1(\tau)\,\mathrm{d}\tau + \frac{M}{M^2 - L_1L_2}\cdot\int_0^t V_2(\tau)\,\mathrm{d}\tau = 0, \tag{11a}$$

$$C_2\frac{\mathrm{d}V_2}{\mathrm{d}t} + \frac{V_2}{R_l} - \frac{L_1}{M^2 - L_1L_2}\cdot\int_0^t V_2(\tau)\,\mathrm{d}\tau + \frac{M}{M^2 - L_1L_2}\cdot\int_0^t V_1(\tau)\,\mathrm{d}\tau = 0. \tag{11b}$$

Here, $R_g(t) < 0$ is the negative resistance that supplies the energy required for self-oscillation, while $R_l > 0$ is the dissipative resistance. The two resonators are taken to be identical, with $L_1 = L_2 = L$ and $C_1 = C_2 = C$. Defining the uncoupled resonance frequency $\omega_0 = 1/\sqrt{LC}$, the gain rate $g(t) = -1/[2R_g(t)C]$, the loss rate $\gamma = 1/(2R_lC)$, and the magnetic coupling coefficient

$k = M/\sqrt{L_1 L_2}$ as in Ref.[1], (11) can be rewritten as

$$\underbrace{\frac{1}{2}\cdot\left(\frac{\mathrm{d}V_1}{\mathrm{d}t}-\frac{\omega_0^2}{k^2-1}\cdot\int_0^t V_1(\tau)\,\mathrm{d}\tau\right)}_{\text{term ①}}-g(t)\cdot V_1\cdot+\frac{1}{2}\cdot\underbrace{\frac{k\cdot\omega_0^2}{k^2-1}\cdot\int_0^t V_2(\tau)\,\mathrm{d}\tau}_{\text{term②}}=0, \tag{12a}$$

$$\underbrace{\frac{1}{2}\cdot\left(\frac{\mathrm{d}V_2}{\mathrm{d}t}-\frac{\omega_0^2}{k^2-1}\int_0^t V_2(\tau)\,\mathrm{d}\tau\right)}_{\text{term ③}}+\gamma\cdot V_2+\frac{1}{2}\cdot\underbrace{\frac{k\cdot\omega_0^2}{k^2-1}\cdot\int_0^t V_1(\tau)\,\mathrm{d}\tau}_{\text{term ④}}=0. \tag{12b}$$

Restricting the limitation $\omega \neq 0$, the Fourier transform of the terms marked ① and ③ gives

$$\mathscr{F}\left[\frac{\mathrm{d}V_n}{\mathrm{d}t}-\frac{\omega_0^2}{k^2-1}\int_0^t V_n(\tau)\,\mathrm{d}\tau\right]=\mathrm{i}\omega\left[1-\frac{\omega_0^2}{(1-k^2)\omega^2}\right]\widetilde{V}_n(\omega). \tag{13}$$

For weak coupling, $k \ll 1$, a first-order Taylor expansion about $\omega = \omega_0$ yields

$$\mathrm{i}\omega\left[1-\frac{\omega_0^2}{(1-k^2)\omega^2}\right]\approx\mathrm{i}\left(\omega-\frac{\omega_0^2}{\omega}\right)\approx 2\mathrm{i}(\omega-\omega_0). \tag{14}$$

Inverse Fourier transform then gives the time-domain approximation

$$\frac{\mathrm{d}V_n}{\mathrm{d}t}-\frac{\omega_0^2}{k^2-1}\int_0^t V_n(\tau)\,\mathrm{d}\tau\approx 2\left(\frac{\mathrm{d}V_n}{\mathrm{d}t}-\mathrm{i}\omega_0 V_n\right). \tag{15}$$

For the terms marked ② and ④,

$$\mathscr{F}\left[\frac{k\omega_0^2}{k^2-1}\int_0^t V_n(\tau)\,\mathrm{d}\tau\right]=\frac{\mathrm{i}k\omega_0^2}{(1-k^2)\omega}\widetilde{V}_n(\omega), \tag{16}$$

Under the same assumptions, $k \ll 1$ and $\omega \approx \omega_0$, the prefactor reduces to $\mathrm{i}k\omega_0$, so

$$\frac{k\omega_0^2}{k^2-1}\int_0^t V_n(\tau)\,\mathrm{d}\tau\approx\mathrm{i}k\omega_0 V_n(t). \tag{17}$$

Substituting (15) and (17) into (12) yields

$$-\mathrm{i}\frac{\mathrm{d}}{\mathrm{d}t}\begin{pmatrix}V_1\\V_2\end{pmatrix}=\begin{pmatrix}\omega_0-\mathrm{i}g(t) & -\kappa\\ -\kappa & \omega_0+\mathrm{i}\gamma\end{pmatrix}\cdot\begin{pmatrix}V_1\\V_2\end{pmatrix}, \tag{18}$$

where $\kappa = k\omega_0/2$ denotes the coupling rate.

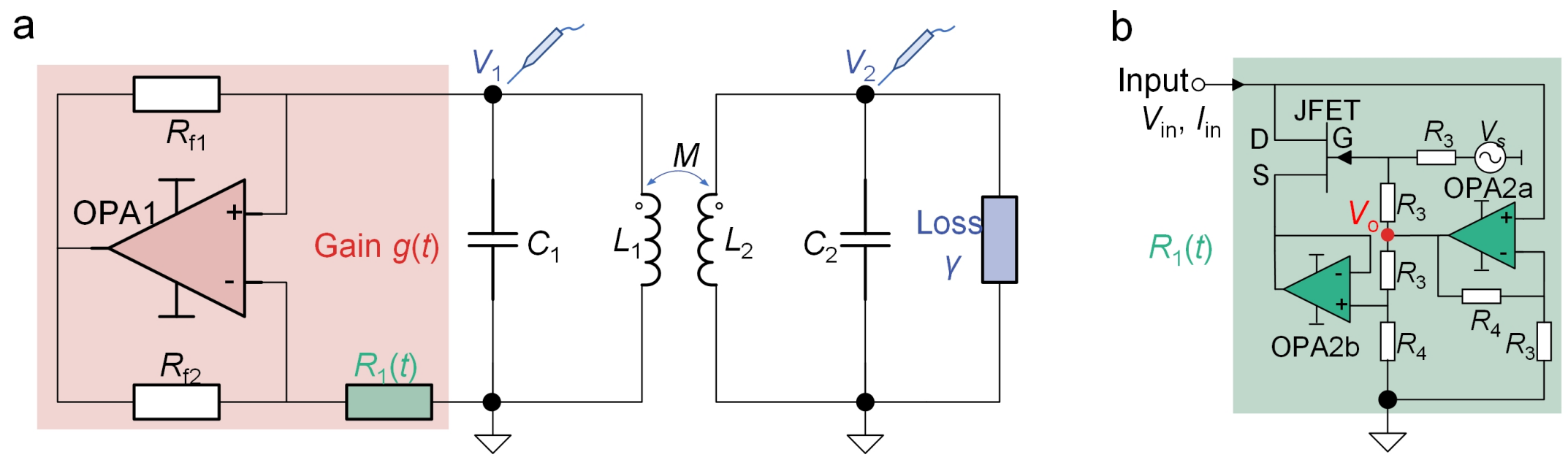


FIG. S3. Circuit schematic of the Floquet $\mathcal{PT}$-symmetric system. (a) Coupled-resonator circuit. (b) Implementation of the dynamic modulation, in which the Floquet-driven gain is introduced through a JFET-based voltage-controlled resistor.

### B. Realization of the Floquet-driven gain

As shown in Fig. S3(a) [reproduced from Fig. 5(a) of the main text], an operational-amplifier negative-resistance stage supplies the energy required for self-oscillation. Its equivalent resistance is $R_g(t) = R_{f1}R_1(t)/R_{f2}$, so periodic variation of $R_1$ modulates the gain,

$$g(t) = -\frac{1}{2C_1 R_g(t)} = \frac{R_{f2}}{2C_1 R_{f1} R_1(t)}. \tag{19}$$

Because $g$ depends inversely on $R_1$, a resistance modulation depth $\alpha_R$ does not translate directly into the gain modulation depth $\alpha$. The gain waveform is therefore obtained from the measured relation $R_1(V_\mathrm{s})$ under the applied voltage $V_\mathrm{s}(t) = V_\mathrm{dc} + V_\mathrm{ac}\cos(\Omega t)$ and is decomposed as

$$\widetilde{g}_n = \frac{1}{T}\int_0^T g(t)\, e^{-\mathrm{i}n\Omega t}\,\mathrm{d}t = \frac{1}{T}\int_0^T \frac{R_{f2}\, e^{-\mathrm{i}n\Omega t}}{2C_1 R_{f1} R_1[V_\mathrm{s}(t)]}\,\mathrm{d}t, \tag{20}$$

where $T = 2\pi/\Omega$. The modulation depth $\alpha$ used in the main text is evaluated as the first-harmonic amplitude relative to the mean gain,

$$\alpha = \frac{2|\widetilde{g}_1|}{\widetilde{g}_0}. \tag{21}$$

In the circuit, $R_1(t)$ is implemented with a JFET voltage-controlled resistor, whose drain–source resistance $R_\mathrm{DS}$ is set by a cosine gate voltage $V_\mathrm{s}$. To improve linearity and extend the usable range of $V_\mathrm{s}$, a clamping network is added [2], as shown in Fig. S3(b). With input voltage $V_\mathrm{in}$, the proportional amplifier OPA2a produces

$$V_\mathrm{o} = \left(1 + \frac{R_4}{R_3}\right)\cdot V_\mathrm{in}. \tag{22}$$

This voltage is divided and buffered by the follower OPA2b, giving the JFET source voltage $V_\mathrm{s}$

$$V_\mathrm{s} = \frac{R_4}{R_3 + R_4} V_\mathrm{o} = \frac{R_4}{R_3} V_\mathrm{in}. \tag{23}$$

The equivalent input resistance $R_1$ is therefore

$$R_1 = \frac{V_\mathrm{in}}{I_\mathrm{in}} = \frac{V_\mathrm{in}}{V_\mathrm{DS}} R_\mathrm{DS} = \frac{V_\mathrm{in}}{V_\mathrm{in} - V_\mathrm{s}} R_\mathrm{DS} = \frac{R_3}{R_3 - R_4} R_\mathrm{DS}. \tag{24}$$

Thus, a periodic modulation of $R_\mathrm{DS}(t)$ produces a proportional modulation of $R_1(t)$, while the ratio $R_3/R_4$ sets the dynamic range of $R_1$.

### C. Simulation methods

The self-oscillation dynamics of the Floquet $\mathcal{PT}$-symmetric circuit were obtained from transient simulations in LTspice, using two complementary models. A device-level model, built from the nominal SPICE models of the JFET and the operational amplifiers, is used to check the theoretical predictions and to assess circuit feasibility at the design stage. JFET characteristics, however, vary between manufacturers and production batches. In the nonlinear resistance regime, even small variations shift the resistance–voltage curve and the modulation voltage required to reach a given $\alpha$. A behavioral model calibrated to the measured resistance characteristics is therefore used for quantitative comparison with experiment and for selecting the operating parameters.

In both models, small initial currents are assigned to the coupled inductors to start the self-oscillation. The coupling sweep is automated with a MATLAB script that updates $\kappa$, launches the LTspice transient run, and holds all other circuit parameters fixed. After a stable self-oscillation state has been established, the voltage waveforms are recorded and their spectral components are extracted by fast Fourier transform.

The device-level schematic is shown in Fig. S4(a). The negative-resistance gain and its temporal modulation are implemented with LT1813 operational amplifiers and a VCR11N JFET; component values are marked on the schematics, and the nonlinear response and saturation of the active devices are retained through their SPICE models. As shown in Fig. S4(b), the simulated self-oscillation frequencies agree with the coupled-mode theory for $f_0 = 191.6$ kHz, $\gamma = 0.092$, $\Omega = 0.1 \times 2\pi f_0$, and $\alpha = 0.3$.

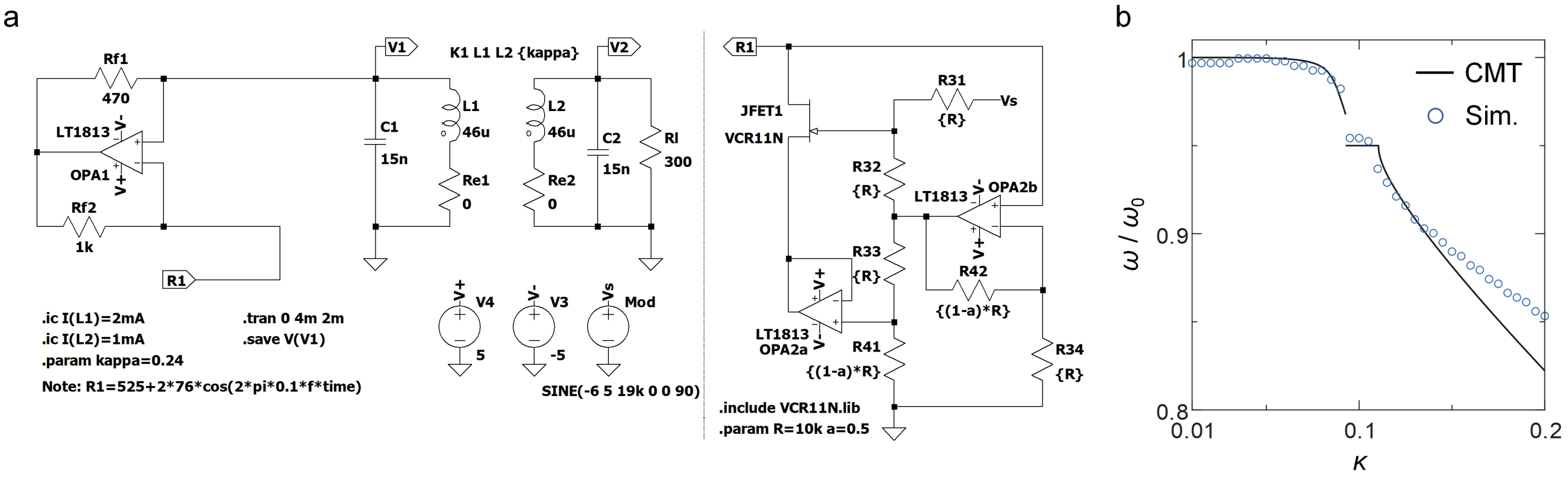


FIG. S4. Device-level simulation of the Floquet $\mathcal{PT}$-symmetric circuit. (a) LTspice schematic based on the SPICE models of the JFET and operational amplifiers. (b) Normalized self-oscillation frequency vesus coupling coefficient $\kappa$, from the device-level simulation (circles) and coupled-mode theory (solid line).

The behavioral model in Fig. S5 is used for the simulation data in Fig. 5(f) of the main text. The JFET resistance module is replaced by a voltage-controlled behavioral resistor whose characteristic is taken from the experiment. The measured resistance–voltage data in Fig. 5(d) of the main text are fitted by

$$R_1(V_s) = Ae^{B \cdot V_s} + Ce^{D \cdot V_s}, \tag{25}$$

with $A = 122.64\ \Omega$, $B = -1.5022\ \mathrm{V}^{-1}$, $C = 596.77\ \Omega$, and $D = -0.1252\ \mathrm{V}^{-1}$. Under sinusoidal modulation, this fit maps the control voltage to the time-varying resistance,

$$V_s(t) = V_{dc} + V_{ac}\cos(\Omega t + \phi), \qquad R_1(t) = R_1[V_s(t)]. \tag{26}$$

Because the mapping is nonlinear, a sinusoidal control voltage does not produce a purely sinusoidal resistance modulation, and the resulting high harmonics can cause deviations from the ideal single-tone Floquet model.

Unlike the device-level model in Fig. S5, which uses nominal component values for theoretical checks and preliminary design, the behavioral model is parameterized from the assembled circuit. The series resistances of the coupled coils are included, and the loss resistance is adjusted for the additional dissipation of the remaining components. In particular, the effective resistance $R_1(t)$ is taken from the measured $R_1(V_s)$ curve rather than from the nominal JFET SPICE model. These differences account for the distinct parameter values used in the two simulations.

Because the attainable range of $R_1(t)$ is highly sensitive to the JFET characteristic, the assembled time-varying-resistance module should be measured before an experiment-specific circuit model is built. The measured resistance–voltage relation then sets the required modulation voltage and is used to tune the remaining circuit parameters.

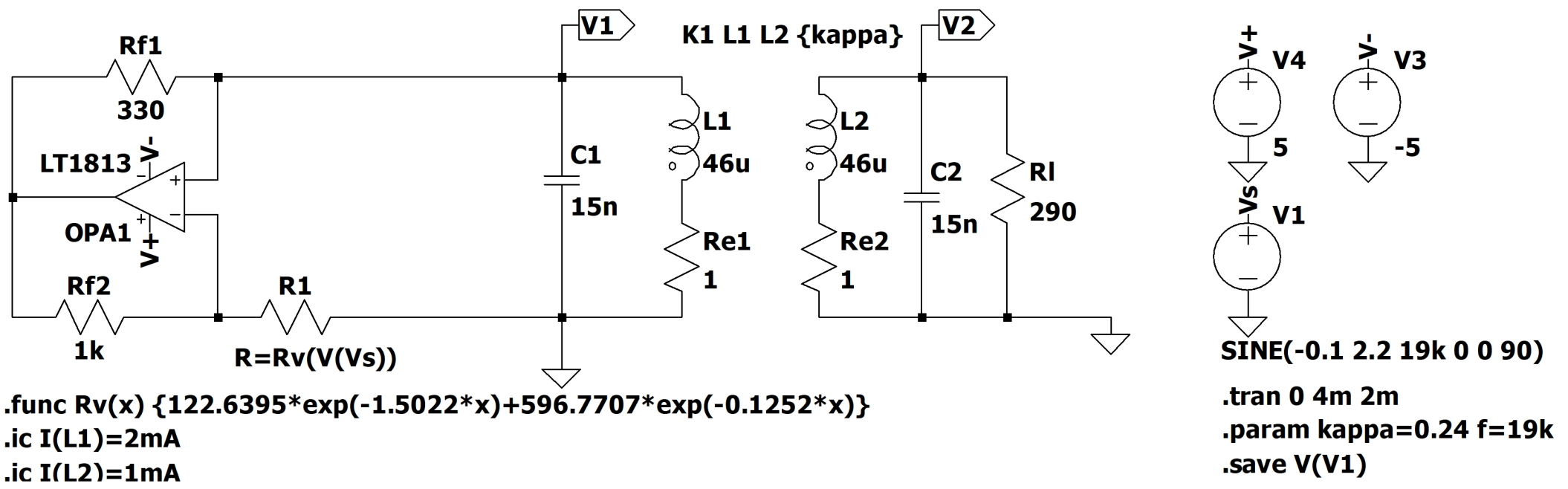


FIG. S5. Behavioral LTspice model used for the data in Fig. 5(f) of the main text. The time-varying resistance follows the measured relation $R_1(V_s)$.

## IV. EXPERIMENTAL SETUP AND MEASUREMENT DETAILS

### A. Measurement of the magnetic coupling coefficient

Fig. S6(a) shows the measured magnetic coupling coefficients $k$ as a function of coil separation $d$. The fitted curve was then used to set the desired $\kappa$ by adjusting the coil separation during the self-oscillation measurements. The coefficient $k$ was obtained with a Keysight E5063A vector network analyzer (VNA). Two identical Würth Elektronik WE-WPCC wireless-power-transfer receiver coils (part number 760308101303) were connected to the analyzer ports through SMA connectors, as shown in Fig. S6(b). The coils were held parallel, with their lateral alignment fixed by the reference line, and $d$ was varied from 4 to 16 mm.

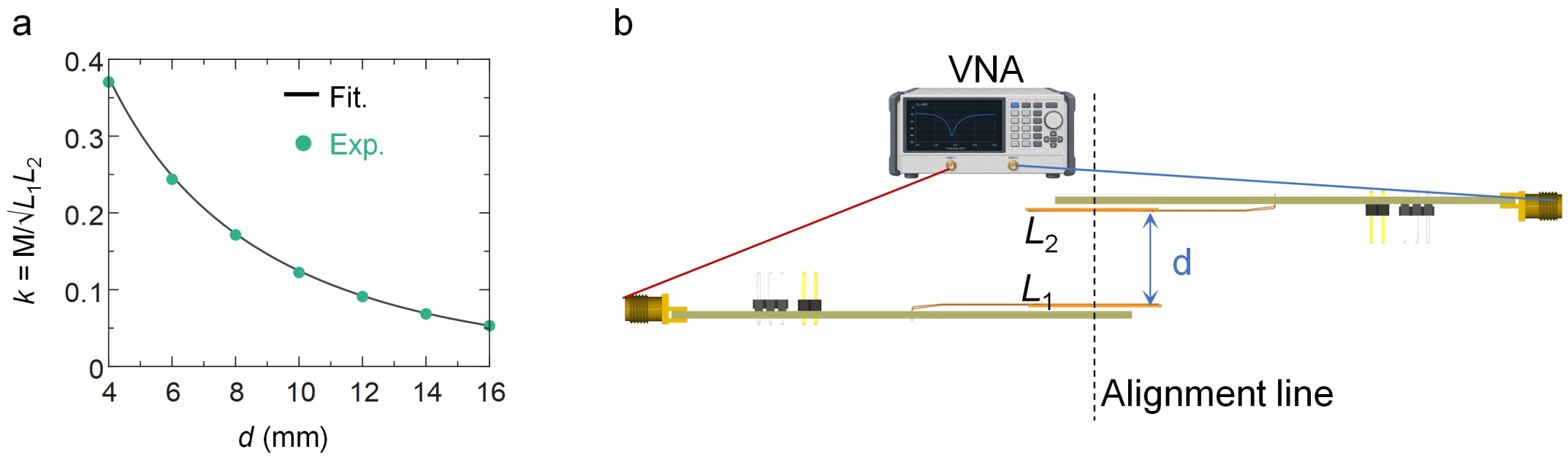


FIG. S6. Measurement of the magnetic coupling coefficient. (a) Measured (circles) and fitted (solid line) values of $k = M/\sqrt{L_1 L_2}$ versus coil separation $d$. (b) VNA setup; the dashed line marks the alignment reference. The coil inductances are $L_1 \approx L_2 \approx 46$ μH.

### B. Hardware implementation

The platform comprises separate gain, modulation, coil, and loss boards, as shown in Fig. S7. Each module is a standard two-layer FR-4 printed circuit board. The circuit is powered by a GW Instek GPP-3323 DC supply, and the modulation signal is provided by a Keysight 33500B waveform generator. The resonator voltage $V_1$ is recorded on a Keysight DSOX4024A oscilloscope. The assembly is mounted on a vertical translation stage so that the coil separation, and hence $\kappa$, can be varied without disturbing the coil alignment. The gain board supplies the negative resistance for the active resonator. An LT1813 operational amplifier implements the negative-impedance converter, and a VCR11N JFET provides the time-varying resistance. The board is driven by $V_s(t)$ from the modulation board and is connected to one coil board to form the gain resonator; $V_1(t)$ is taken from its measurement port.

The modulation board is an OPA2192 operational summing amplifier. It combines the sinusoidal output of the Keysight 33500B waveform generatorwith an adjustable DC reference and produces $V_s(t) = V_{dc} + V_{ac}\cos(\Omega t)$, which sets the gate voltage of the VCR11N JFET on the gain board. The two coil boards carry identical Würth Elektronik WE-WPCC wireless-power-transfer receiver coils (part number 760308101303), which form the inductive elements of the resonators. One coil board is connected to the gain board and the other to the loss board, giving the active and lossy resonators; the resonators interact through the magnetic coupling of the coils. The loss board sets the passive dissipation of the lossy resonator. It contains the loss resistance, connects to the second coil board, and provides the measurement port for $V_2(t)$. The component values used for the experimentin Fig. 5(f) of the main text are listed in Table S1.

---

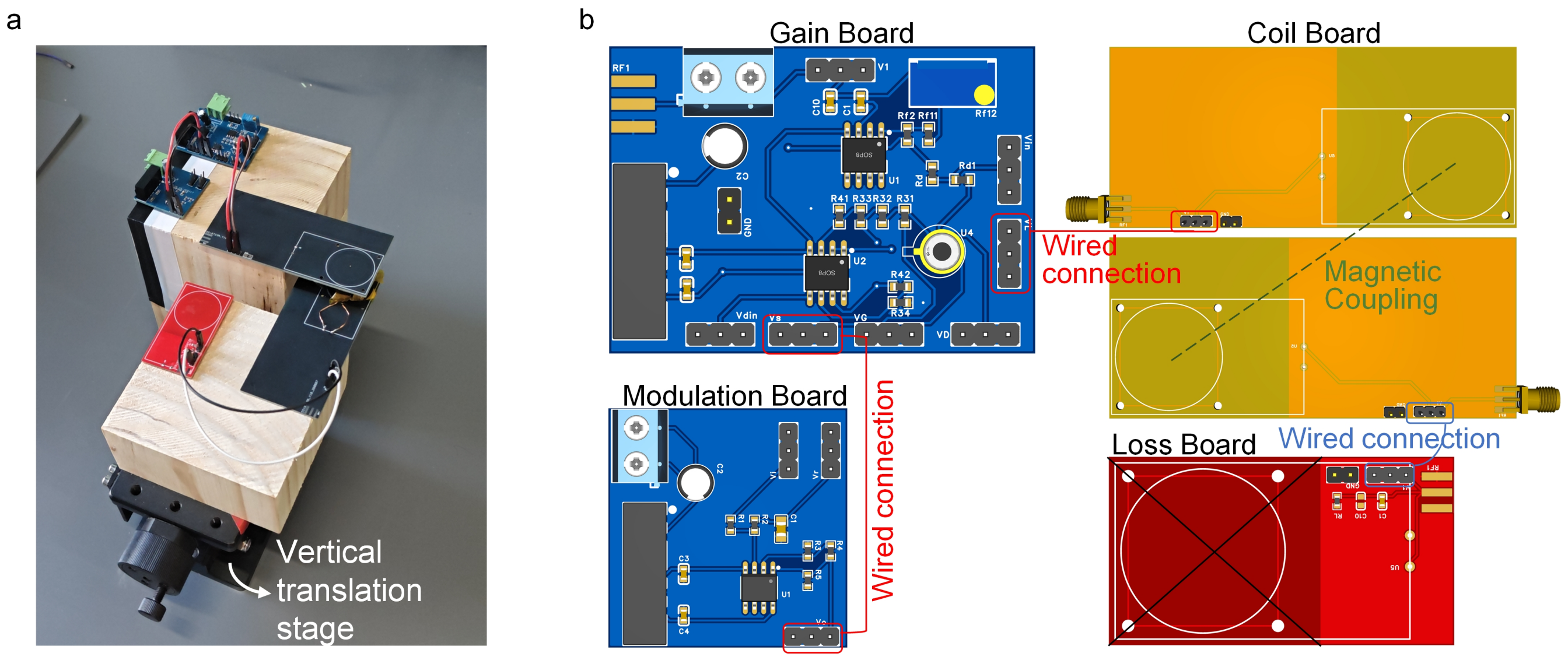


FIG. S7. Hardware implementation of the Floquet $\mathcal{PT}$-symmetric circuit. (a) Assembled circuit on the vertical translation stage used to set the coil separation. (b) Modular layout and interconnections of the gain, modulation, coil, and loss boards. The resonators are coupled magnetically through the coil boards.

TABLE S1. Component values used for the experiment in Fig. 5(f) of the main text.

| **Main circuit** | | |
|---|---|---|
| Parameter | Description | Value |
| $L_1$, $L_2$ | Receiver coil (WE-WPCC 760308101303) | 46 μH |
| $C_1$, $C_2$ | Resonant capacitances | 15 nF |
| $R_{f1}$ | Negative-impedance feedback resistance | 220 Ω |
| $R_{f2}$ | Negative-impedance feedback resistance | 1 kΩ |
| $R_l$ | Loss resistance | 300 Ω |
| OPA1 | Negative-impedance converter | LT1813 |
| $f_0$ | Uncoupled resonance frequency | ≈ 191.6 kHz |
| **Time-varying resistance** | | |
| JFET | Voltage-controlled resistor | VCR11N |
| $R_{31}$, $R_{32}$, $R_{33}$, $R_{34}$ | Amplifier resistances | 10 kΩ |
| $R_{41}$, $R_{42}$ | Divider resistances | 5 kΩ |
| OPA2a, OPA2b | Modulation operational amplifiers | LT1813 |
| $\Omega/(2\pi)$ | Modulation frequency | 19 kHz |
| $V_s(t)$ | JFET modulate voltage | $-0.1+2\cos(\Omega t)$ V |



\end{document}